\documentclass[
     ,final            
    ]
    {aipproc}

\layoutstyle{6x9}

\begin{document}

\def\mib#1{\mbox{\boldmath $#1$}}

\title{Radiative Corrections to Low-Energy Neutrino-Deuteron Reactions
 Revisited}


\author{Takahiro Kubota}{
  address={Department of Physics, Graduate School of Science, 
\\
Osaka University, Toyonaka, 
  Osaka 560-0043, Japan}
}

\begin{abstract}
The one-loop QED and electroweak radiative corrections to 
neutrino-deuteron scattering induced by the neutral current are 
reexamined, paying a particular attention to the constant term which 
has never been treated properly in literature. 
This problem is closely related to the definition of 
the axial-vector coupling constant $g_{A}$ 
and requires thorough calculations 
of the constant terms in the charged current processes, too. 
We find that the radiative corrections to the neutral current induced 
reactions amount to 1.7  (1.5) per cent enhancement, if  
the Higgs boson mass is $m_{H}=1.5 \: m_{Z} (m_{H}=5.0 \: m_{Z})$
This number happens to be close to  that given by 
Kurylov et al., but we argue that this is  accidental.
\end{abstract}

\maketitle



This talk is based on the collaboration with Masataka Fukugita 
\cite{fukugita1} (see also \cite{fukugita2, fukugita3}), and is 
concerned with radiative 
corrections to the reactions occurring at   
the Sudbury Neutrino Observatory (SNO).
The observation of the solar neutrinos at SNO  
has been playing  important roles to resolve the solar neutrino 
problem \cite{bahcall}.
The measurement of neutrino-deuteron scattering 
\begin{eqnarray}
\nu _{e}+d &\longrightarrow & e^{-}+p+p, 
\label{eq:cc}
\\
\nu _{e}+d &\longrightarrow & \nu _{e}+p+n
\label{eq:nc}
\end{eqnarray}
has now reached the level that  radiative corrections should be 
included in the analyses  \cite{sno}.
The first step  toward  evaluation of  the radiative corrections to 
(\ref{eq:cc}) and (\ref{eq:nc}) was taken by Towner \cite{towner}. 
Some subtle problems associated with soft photon emission were pointed out 
\cite{beacom} and have been  solved \cite {kurylov} by giving due 
consideration to the energy-dependence of the wave function overlap 
between initial and final states. 
There has remained, however, the problem as to the constant terms of  
the radiative correction, as remarked by the authors of  \cite{kurylov}.
They have evaluated the corrections to (\ref{eq:cc}) by assuming 
implicitly that the 
inner correction to the Gamow-Teller part is the same as that to the 
Fermi transition.


The transition amplitudes squared of the charged current processes 
in general are expressed in general on the ${\cal O}(\alpha )$ level  
as 
\begin{eqnarray}
A(\beta) =\left (1+\delta_{\rm out}(\beta) \right )\left[
f_{V}^{2} \left (1+{\delta_{\rm in}^{\rm F}}
\right ) \langle 1 \rangle ^{2} +
g_{A}^{2} \left (1+
{\delta_{\rm in}^{\rm GT}}\right )\langle {\mib \sigma} \rangle ^{2}
\right]\ , 
\label{eq:Aterm}
\end{eqnarray}
where $f_{V}(\equiv 1)$ and $g_{A}$ are the vector and axial-vector 
coupling constants. 
The Fermi and Gamow-Teller matrix elements are denoted by 
$\langle 1 \rangle $ and $\langle {\mib \sigma} \rangle$, respectively.
The outer correction $\delta _{\rm out}(\beta )$ is a function of the 
electron velocity $\beta $ and is  process-dependent \cite{ks}.
 (See also ref. \cite{ando1} for an approach based on effective 
field theory.) 
The inner corrections $\delta _{\rm in}^{\rm F}$ and 
$\delta _{\rm in}^{\rm GT}$ are in contrast independent of 
charged current processes considered and are universal constants 
\cite{sirlin, fukugita2}.
The Fermi part inner correction $\delta _{\rm in}^{\rm F}$ 
has been known for long time \cite{ms1}, 
while the Gamow-Teller part $\delta _{\rm in}^{\rm GT}$ was calculated 
only recently \cite{fukugita2}. 


The axial-vector coupling constant $\tilde g_{A}$ usually quoted in literature 
is extracted from the neutron beta decay using the formula
\begin{eqnarray}
A(\beta) =\left (1+\delta_{\rm out}(\beta )\right )\left (1+{\delta_{\rm
in}^{\rm F}}
\right ) \left[ \langle 1 \rangle ^{2}
f_{V}^{2}
+\langle {\mib \sigma} \rangle ^{2}
\tilde g_{A}^{2} \right]\ .
\label{eq:Aterm1}
\end{eqnarray}
The polarized neutron beta decay is also used to extract   
$\tilde g_{A}$ instead of  $g_{A}$ \cite{fukugita3}. 
The relation between $g_{A}$ and $\tilde g_{A}$ is obviously given by 
\begin{eqnarray}
g_{A}^{2}= \left ( \frac{1+{\delta_{\rm in}^{\rm F}}}
{1+{\delta_{\rm in}^{\rm GT}}}\right ) ~ \tilde g_{A}^{2}\ ,
\label{eq:pseudoga}
\end{eqnarray}
and this indicates  $g_{A}\neq \tilde g_{A}$, 
because of  $\delta _{\rm in}^{\rm F}\neq \delta _{\rm in}^{\rm GT}$ 
as was shown by explicit computation in \cite{fukugita2}. 
This, however, does not cause 
any practical problem, since the relation between the ``bare" $g_{A}$ and 
the ``redefined " $\tilde g_{A}$ is universal, so far as we consider 
only charged current processes \cite{sirlin, gudkov}. 
Thus the analysis of (\ref{eq:cc}) 
in \cite{sno} need not be corrected essentially.


If we consider neutral current processes such as (\ref{eq:nc}),
however,  we have to be more careful about the finite radiative corrections to 
 the axial-vector coupling constant. The reaction (\ref{eq:nc}) 
is purely of the Gamow-Teller type and its amplitude squared is written as 
\begin{eqnarray}
B(\beta)= (1+ \Delta_{\rm in}^{\rm GT})g_{A}^{2} \langle {\mib \sigma}
\rangle ^{2} \ , 
\label{eq:corrNC}
\end{eqnarray}
where $\Delta _{\rm in}^{\rm GT}$ is the  ${\cal O}(\alpha )$ radiative 
corrections. Using the relation (\ref{eq:pseudoga}),  
we see that (\ref{eq:corrNC}) is expressed in terms of $\tilde g_{A}$ as 
\begin{eqnarray}
B(\beta)= (1+ \Delta_{\rm in}^{\rm GT})
\left ( \frac{1+\delta _{\rm in}^{\rm F}}{1+\delta _{\rm in}^{\rm GT}}\right )
{\tilde g}_{A}^{2} \langle {\mib \sigma}
\rangle ^{2} \ . 
\end{eqnarray}


In \cite{fukugita1}, we have extracted $\Delta _{\rm in}^{\rm GT}$ 
on the bases of the work of Marciano and Sirlin \cite{ms2} and have found 
\begin{eqnarray}
\Delta _{\rm in}^{\rm GT}&=&0.0192 \hskip1cm {\rm for} \:\: m_{H}=1.5 \:
m_{Z}, 
\\
\Delta _{\rm in}^{\rm GT}&=&0.0173 \hskip1cm {\rm for} \:\: m_{H}=5.0 \:
m_{Z}.
\end{eqnarray}
Combining these results with the previous calculation of the inner 
corrections \cite{fukugita2} 
\begin{eqnarray}
\delta _{\rm in}^{\rm F}=0.0237, \hskip1cm \delta _{\rm in}^{\rm GT}=
0.0262, 
\end{eqnarray}
we find that the cross section of (\ref{eq:nc}) is enhanced by the factor 
\begin{eqnarray}
\left (1+\Delta _{\rm in}^{\rm GT} \right ) \left (
\frac{1+\delta _{\rm in}^{\rm F}}{1+\delta _{\rm in}^{\rm GT}}\right )
=1.017 & & \hskip1cm {\rm for} \:\: m_{H}=1.5 \:m_{Z}, 
\nonumber \\
\left (1+\Delta _{\rm in}^{\rm GT} \right ) \left (
\frac{1+\delta _{\rm in}^{\rm F}}{1+\delta _{\rm in}^{\rm GT}}\right )
=1.015  & & \hskip1cm {\rm for} \:\: m_{H}=5.0 \:m_{Z}  \ .
\label{eq:1017}
\end{eqnarray}
Note that this is rather close to the number given in  \cite{kurylov} 
and that the analysis of (\ref{eq:nc}) in  \cite{sno} using \cite{kurylov} 
need not be altered basically. We should, however, like to emphasize that the 
approximate  agreement of our results (\ref{eq:1017}) with that 
of  \cite{kurylov} is simply due to an accidental  cancellation 
of errors of the latter,  between those caused by putative identification 
of constant terms for the Fermi and Gamow-Teller transitions for the 
charged current reactions and minor errors in their treatment of the 
constant terms for neutral current induced reactions. 

Throughout the present work, the so-called one-body impulse approximation 
is used and the effect of the spectator nucleon is not included. 
See ref. \cite{ando2} in this connection for an approach based on heavy-baryon 
chiral perturbation theory.  
We note as a final remark that the constant term for
the radiative correction
to the ratio of neutral to charged current reaction (after the usual outer
correction \cite{kurylov, ks} for the charged 
current reaction ) is $-0.6$\%, which may
be compared with the claimed error (0.5\%) of nuclear calculations for the
ratio of tree level cross sections \cite{nakamura}.





\begin{theacknowledgments}
The author would like to express his sincere thanks to Professor 
Masataka Fukugita for stimulating discussions, fruitful collaboration 
and a careful reading of this manuscript. This work is supported 
in part by the Grant in Aid from the Ministry of Education and JSPS.
\end{theacknowledgments}



\bibliographystyle{aipprocl} 

\bibliography{sample}

\begin{thebibliography}{9}
\bibitem{fukugita1}
M. Fukugita and T. Kubota,  {\it Phys. Rev.} {\bf D 72}, 071301 (2005).
%
\bibitem{fukugita2}
M. Fukugita and T. Kubota,  {\it Acta Physica Polonica}  
{\bf B 35}, 1687 (2004). 

\bibitem{fukugita3}
M. Fukugita and T. Kubota,  {\it Phys. Lett.} {\bf B 598}, 67 (2004).
%
%
\bibitem{bahcall}
J.N. Bahcall, ``{\it Neutrino Astrophysics}" (Cambridge 
University Press, 1989).
%
\bibitem{sno}
B. Aharmim et al. (SNO Collaboration), arXiv:nucl-ex/0502021.
%
%
\bibitem{towner}
I.S. Towner,  {\it Phys. Rev. }  {\bf C 58}, 1288 (1998).
%
\bibitem{beacom}
J.F. Beacom and S.J. Parke,  {\it Phys. Rev.}  {\bf D 64}, 091302 (2001).
%
\bibitem{kurylov}
A. Kurylov, M.J. Ramsey-Musolf and P. Vogel,  {\it Phys. Rev. }
 {\bf C 65}, 055501 (2002);  {\bf C 67}, 035502 (2003).
%
\bibitem{ks}
T. Kinoshita and A. Sirlin,
 {\it Phys. Rev.}  {\bf 113}, 1652 (1959);
P. Vogel,  {\it Phys. Rev.}   {\bf D 29}, 1918 (1984);
S.A. Fayans,  {\it Yad. Fiz.} {\bf 42}, 929 (1985)  
[ {\it Sov. J. Nucl. Phys.} {\bf 42}, 590 (1985)].
%
\bibitem{ando1}
S. Ando et al., {\it Phys. Lett.} {\bf B 595}, 250 (2004).
%
\bibitem{sirlin}
A. Sirlin,  {\it Phys. Rev.}  {\bf 164}, 1767 (1967);
see also E.S. Abers, D.A. Dicus, R.E. Norton and H.R. Quinn,  
{\it Phys. Rev.} {\bf 167}, 1461 (1968).
%
\bibitem{ms1}
W.J. Marciano and A. Sirlin,  {\it Phys. Rev. Lett.}  
{\bf 56}, 22 (1986).
%
\bibitem{gudkov}
V. Gudkov, J. Neutron Res. {\bf 13}, 39 (2005); 
V. Gudkov et al., J. Res. Natl. Inst. Stand. Technol. {\bf 110}, 315 (2005).  
%
\bibitem{ms2} W.J. Marciano and A. Sirlin,  {\it Phys. Rev.}    
{\bf D 22}, 2695 (1980).
%
\bibitem{ando2}
S. Ando et al., {\it Phys. Lett.} {\bf B 555}, 49 (2003). 
%
\bibitem{nakamura}
S. Nakamura, T. Sato, V. Gudkov and K. Kubodera,  {\it Phys. Rev.}  
 {\bf C 63}, 034617 (2001); S. Nakamura et al., {\it Nucl. Phys.} 
{\bf A 707}, 561 (2002).
%
\end{thebibliography}

\IfFileExists{\jobname.bbl}{}
 {\typeout{}
  \typeout{******************************************}
  \typeout{** Please run "bibtex \jobname" to optain}
  \typeout{** the bibliography and then re-run LaTeX}
  \typeout{** twice to fix the references!}
  \typeout{******************************************}
  \typeout{}
 }



\end{document}